\documentclass[aps,pra,preprint]{revtex4-1}

\usepackage{graphicx}
\usepackage{amsmath}

\newcommand{\func}[2]{#1\!\left[#2\right]}
\newcommand{\funcSq}[2]{#1\!\left[#2\right]}
\newcommand{\tcav}{t_c}
\newcommand{\sinc}{\mathrm{sinc}}
\newcommand{\dP}{\Delta\!P}
\newcommand{\dPhat}{\Delta\!\widehat{P}}
\newcommand{\Delom}{\Delta\omega}
\newcommand{\abs}[1]{\left|#1\right|}
\newcommand{\ket}[1]{\left|#1\right\rangle}
\newcommand{\un}[2]{#1\,\mathrm{#2}}
\newcommand{\rf}{rf }
\newcommand{\trans}[2]{\mbox{[#1 to #2]}}
\newcommand{\TE}{$\mathrm{TE}_\mathrm{011}$ }

\newcommand{\sref}[1]{section~\ref{#1}}

\newcommand{\eref}[1]{Eq.~(\ref{#1})}

\newcommand{\fref}[1]{Fig.~\ref{#1}}
\newcommand{\Fref}[1]{Fig.~\ref{#1}}
\newcommand{\aref}[1]{appendix~\ref{#1}}

\newcommand{\Yfunc}{G} 
\newcommand{\Xfunc}{U} 

\begin{document}


\title{Atomic trajectory characterization in a fountain clock based on the spectrum of a hyperfine transition} 



\author{N. Nemitz}
\email[]{Nils.Nemitz@PTB.de}

\author{V. Gerginov}
\author{R. Wynands}
\author{S. Weyers}
\affiliation{Physikalisch-Technische Bundesanstalt (PTB), Bundesallee 100, 38116 Braunschweig, Germany}


\date{\today}

\begin{abstract}
We describe a new method to determine the position of the atomic cloud during its interaction with the microwave field in the cavity of a fountain clock. The positional information is extracted from the spectrum of the \mbox{$\ket{F=3, m_F=0}$ to $\ket{F=4, m_F=-1}$} hyperfine transition, which shows a position dependent asymmetry when the magnetic C-field is tilted by a few degrees with respect to the cavity axis. Analysis of this spectral asymmetry provides the horizontal center-of-mass position for the ensemble of atoms contributing to frequency measurements. With an uncertainty on the order of $\un{0.1}{mm}$, the obtained information is useful for putting limits on the systematic uncertainty due to distributed cavity phase gradients. The validity of the new method is demonstrated through experimental evidence.
\end{abstract}


\maketitle 

\section{Introduction}
Caesium fountain clocks provide the best available implementation of the SI second with a relative uncertainty below $10^{-15}$~\cite{Wynands2005}. As such they are crucial to the accuracy of International Atomic Time (TAI). 


The basic experimental setup of a fountain clock is well established: A large number of atoms is laser-cooled to $\mathrm{\mu K}$ temperatures. The atomic cloud is then launched vertically, passing through a cavity during its upwards motion. Here it is exposed to a microwave field that drives the clock transition and puts the atoms into a quantum mechanical superposition state. Gravity causes the atoms to slow down and fall back, where they cross the same cavity again and experience a second microwave pulse. This sequence implements Ramsey's method of separated oscillatory fields with a single cavity, avoiding any frequency shift resulting from a global phase difference between the two field regions. However, the atoms generally traverse the cavity at different horizontal positions during the upwards and the downwards passage, such that position dependent phase variations of the microwave field cause a frequency bias. This gives rise to one of the leading contributions to the uncertainty budget of several atomic fountain clocks~\cite{Wynands2005}. An improved understanding of the actual microwave fields and phase gradients~\cite{Li2004,Li2010} has recently led to a renewed investigation of these effects~\cite{Guena2011PRL,Szymaniec2011,Weyers2012} and it has been shown that under normal operating conditions the largest frequency shifts can be expected to occur due to phase gradients oriented horizontally across the cavity. Such gradients result from imbalances in the microwave feeds or from inhomogeneities of the electrical conductivity of the cavity material.

An assessment of the resulting contribution to the frequency uncertainty requires information on the horizontal position of the cloud during its upwards and downwards cavity passage. So far, no method for an in-situ measurement has been available. Estimates from images of the atom cloud before the launch, or possibly during the passages through the detection zone, carry a particular difficulty: Information is needed on the distribution of those atoms that are detected at the end of the sequence, and this can differ significantly from the full distribution of all launched atoms. The full atomic trajectories and their interaction with the known obstructions inside the fountain can be taken into account by simulations, but the results strongly depend on the assumed cloud parameters.

For the majority of cavity designs in use, it is possible to deliberately apply a phase gradient in the cavity by asymmetric feeding. This allows an adjustment of the launch direction to effectively cancel the shift of the cloud position between passages for one coordinate direction~\cite{Guena2011PRL,Szymaniec2011}. Improved cavity designs would permit this for both directions~\cite{Li2010}, but have not been implemented in any fountain yet. Furthermore, this method requires manual adjustment of the fountain alignment and considerable measurement time, making it unsuitable for quick confirmation measurements or even on-line monitoring of the cloud position. The cavity design used in the PTB fountains~\cite{Schroeder2002} favors stability of the microwave feeding over adjustability and does not provide this option at all.

In this paper, we present a new method to find the center-of-mass position during the upwards and the downwards cavity passages for the same fraction of atoms in the cloud that also contributes to the frequency measurement. The method is based on position dependent asymmetries in the transition probability spectra of the $\ket{F=3, m_F=0}$ to $\ket{F=4, m_F=-1}$ transition. Its use in~\cite{Weyers2012} has helped reduce the uncertainty of the frequency bias from the cavity phase distribution in PTB's fountain CSF2 to $1.33 \times 10^{-16}$ and the total systematic uncertainty to $u_B = 4.1 \times 10^{-16}$.

Section~\ref{sec_field_interaction} of this paper aims to provide an illustrative explanation for the asymmetry of the measured spectra and its relation to the magnetic field in the cavity. In section~\ref{sec_analytical_model} these considerations are developed into an approximated analytical model. This provides a simple way to determine the orientation of the magnetic field and to extract the center-of-mass position for the ensemble of atoms contributing to frequency measurements. The results of these calculations are consistent with experimental data, as presented in \sref{sec_experimental_test_main}. The uncertainty of the position determination is discussed in \sref{sec_errors}.


\section{Atom-field interaction for $\Delta m = \pm 1$ transitions}
\label{sec_field_interaction}

In the fountain CSF2, the atoms are state-selected for the $\ket{F=3, m_F=0}$ state by a combination of microwave interaction and light pressure  before entering the cavity~\cite{Gerginov2010}. A nominally vertical magnetic field, usually referred to as the ``C-field'', makes the transitions to different Zeeman sublevels of the $\ket{F=4}$ hyperfine state individually addressable.  We will use the notation \trans{$a$}{$b$} to refer to transitions $\ket{F=3, m_F=a}$ to $\ket{F=4, m_F=b}$. 


\subsection{Field distribution in a fountain clock cavity}
\label{sec_field_distribution}

Fountain clocks use cylindrical cavities designed such that the \TE mode is resonant at the frequency of the \trans{0}{0} clock transition. The atoms enter and leave the cavity through circular apertures in the top and bottom endcaps. The diameter of these apertures is chosen below the wavelength of the microwave signals such that the field distribution remains confined. The addition of cutoff tubes of the same diameter further reduces microwave leakage, which might otherwise lead to a bias of the measured clock frequency~\cite{Weyers2006,Shirley2006}.


The actual fields in the cavity, including the effects of endcap holes and wall losses, can be calculated using finite element methods~\cite{Li2004,Li2010}. For an analytic description, we will approximate the field distribution by that in a perfectly conducting cavity without holes. For a cavity with height $d$ centered around the origin, this is described in cylindrical coordinates by the equations
\begin{eqnarray}
\label{eq_amp_Hrho}
&H_\rho
  & = \dfrac{\pi^2}{2 \, \gamma \, d^2} \,       
      \func{J_1}{ \gamma \rho }
      \func{\sin}{ \dfrac{\pi \, z}{d} }      
\\
&H_z
\label{eq_amp_Hz}
  & = \dfrac{\pi}{2 \, d} \, 
      \func{J_0}{ \gamma \rho } 
      \func{\cos}{ \dfrac{\pi \, z}{d} }
      \ .
\end{eqnarray}
The Bessel functions $J_0$ and $J_1$ describe the field dependence on the radial coordinate $\rho$, and  $\gamma = x_0 / R$, where $x_0$ is the first zero of $J_1$ and $R$ is the cavity radius. In the PTB fountains CSF1 and CSF2, $R=\un{24.2}{mm}$ and $d=\un{28.12}{mm}$~\cite{Schroeder2002}. We will take the equations to describe the instantaneous field distribution at $t=0$, and they have been normalized such that at $\rho=0$ the integral of $H_z$ over the vertical coordinate $z$ is
\begin{equation}
  \int_{-d/2}^{d/2}{ \func{H_z}{\rho=0} \, dz } = 1
  \ .
\end{equation}

The approximation reproduces the key features of the field distribution very well: When traveling vertically through the cavity, the sign of the horizontal field amplitudes $H_x$ and $H_y$ reverses at the center plane. Moving radially, $H_x$ and $H_y$ are zero on the cavity axis and increase with radial position $\rho$. In the inner region that is accessible to atoms traveling through the endcap apertures, they are generally small compared to the vertical amplitude $H_z$, except at the top and bottom, where $H_z$ becomes small. Radially, $H_z$ has its largest value on the cavity axis and falls off slightly with increasing $\rho$.

An atom is assumed to enter the cavity at $-\tcav /2$ and leave it again at $+\tcav /2$, with $\tcav=\un{10.6}{ms}$ for CSF2. During the traversal, the motion of the atoms translates the spatial distribution of the field into a temporal pulse. Due to the extremely narrow linewidth of the clock transition, the amplitude and phase of this pulse determine the features of the transition probability spectrum.


\subsection{Angle -- phase equivalence}
\label{sec_angle_and_phase}

The most significant phase effect results from the local direction of the \rf field. An oscillating horizontal field component oriented at an angle $\beta$ to the $x$-axis and with a peak amplitude $H_0$, can be written as a vector in $xy$-coordinates:
%
%
%
\begin{equation}
  \label{eq_linear_field_at_beta}
  \func{\vec{H}_h}{\beta,t} 
  = \underbrace{
   \frac{H_0}{2}
   \left(\!
     \begin{array}{c}
        \ \,  \func{\cos}{ \omega_\mathrm{\rf} \, t - \beta }\\
        \!\! -    \func{\sin}{ \omega_\mathrm{\rf} \, t - \beta }
     \end{array}
  \!\right)
  }_{\func{\vec{H}_\mathrm{cw}}{\beta,t} }
  + \underbrace{
   \frac{H_0}{2}
   \left(\!
     \begin{array}{c}
       \func{\cos}{ \omega_\mathrm{\rf} \, t + \beta }\\
       \func{\sin}{ \omega_\mathrm{\rf} \, t + \beta }
     \end{array}
  \!\right) 
  }_{\func{\vec{H}_\mathrm{ccw}}{\beta,t} }
  .
\end{equation}
Physically, this treats the linearly polarized field as the superposition of two opposite circular polarization contributions $\vec{H}_\mathrm{cw}$ and $\vec{H}_\mathrm{ccw}$. Each of these carries the angular momentum along the nominal C-field axis that is required to drive one of the $\Delta m = \pm 1$ transitions. We have chosen to investigate the \trans{0}{-1} transition, which has the same initial $\ket{F=3,m_F=0}$ state used in clock operation and interacts with the clockwise circular contribution $\vec{H}_\mathrm{cw}$ (e.g., see \cite{Foot2005}). In the relevant term, the angle $\beta$ appears as a phase retardation. In essence, atoms experience any change in the orientation of the linearly polarized \rf field as a change in its phase.

The $\vec{H}_\mathrm{ccw}$ contribution, for which $\beta$ advances the phase instead, can only excite the \trans{0}{1} transition. The Zeeman splitting is usually large enough that the off-resonant excitation of this transition does not add significantly to the transition probability while measuring the spectrum of the \trans{0}{-1} transition.

%


\subsection{Shape of $\Delta m = \pm 1$ spectra}


We now investigate the changes in the direction of the \rf field experienced by an atom crossing the cavity. Let us assume that the reversal of the horizontal \rf component between the top and the bottom of the cavity occurs in a smooth clockwise rotation. Note that this rotation is slow compared to that of $\vec{H}_\mathrm{cw}$, and the field is best described as having a linear polarization with a slowly varying direction $\func{\beta}{t} = - \pi \, t / \tcav$. This slow rotation causes a gradual phase shift of the clockwise contribution driving the \trans{0}{-1} transition, which can be expressed as a shift of the effective \rf frequency by $\omega_\mathrm{rot} = d \beta / dt = \pi / \tcav$.
The appearance of frequency shifts for the $\Delta m = \pm 1$ transitions when the direction of the microwave field changes relative to that of the static magnetic field is known as the Millman effect~\cite{Millman1939,Vanier1984}.

However, the horizontal field component normally does not rotate during the cavity traversal, but simply decreases to zero amplitude at the center plane and then reappears oriented in the opposite direction. This behavior can be modeled by taking the previously assumed linearly polarized field with its slow clockwise rotation and adding a second linearly polarized field with a slow counter-clockwise rotation. This additional field then has its own fast-rotating, circular contributions, with only the clockwise contribution driving the \trans{0}{-1} transition. The opposite direction of the slow rotation reverses the phase shift with time, such that the effective frequency of the second field has an opposite shift of $-\omega_\mathrm{rot}$.

The presence of these two components in the \rf field affecting the \trans{0}{-1} transition suggests a single-passage transition probability spectrum with two peaks near $\pm \omega_\mathrm{rot}$ relative to the unperturbed resonant frequency of the atoms. 
This is quite close to what is observed~\cite{Giordano1995}. In CSF2, where  $\omega_\mathrm{rot} = \pi / \tcav = 2 \pi \times \un{47.2}{Hz}$, the two main lobes appear at $\pm 2\pi \times \un{64.6}{Hz}$ due to interference effects that are not covered in this simple approach.


\subsection{Microwave pulse in the presence of a C-field tilt}

\begin{figure}[t!b]
  \begin{center}
    \includegraphics[width=86 mm]{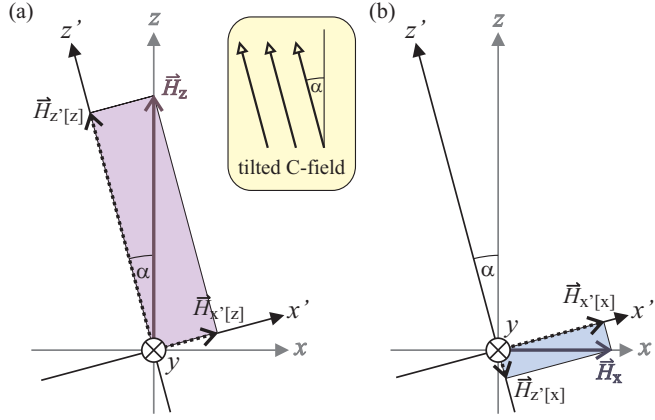}
  \end{center}
  \caption{Mixing of cavity \rf field components: The tilt $\alpha$ of the C-field defines a new coordinates system $(x',y,z')$ (note the shared $y$-axis). In this system, the field component $H_{z'}$, consisting of the projected components $H_{z'[z]}$ (a) and $H_{z'[x]}$ (b), drives the \trans{0}{0} transition (a). \trans{0}{-1} is driven by the unchanged $H_y$ component and $H_{x'}$, consisting of the projections $H_{x'[z]}$ (a) and $H_{x'[x]}$ (b).}
\label{fig_tilted_coordinates}
\end{figure}

\begin{figure}[t!b]
  \begin{center}
    \includegraphics[width=76 mm]{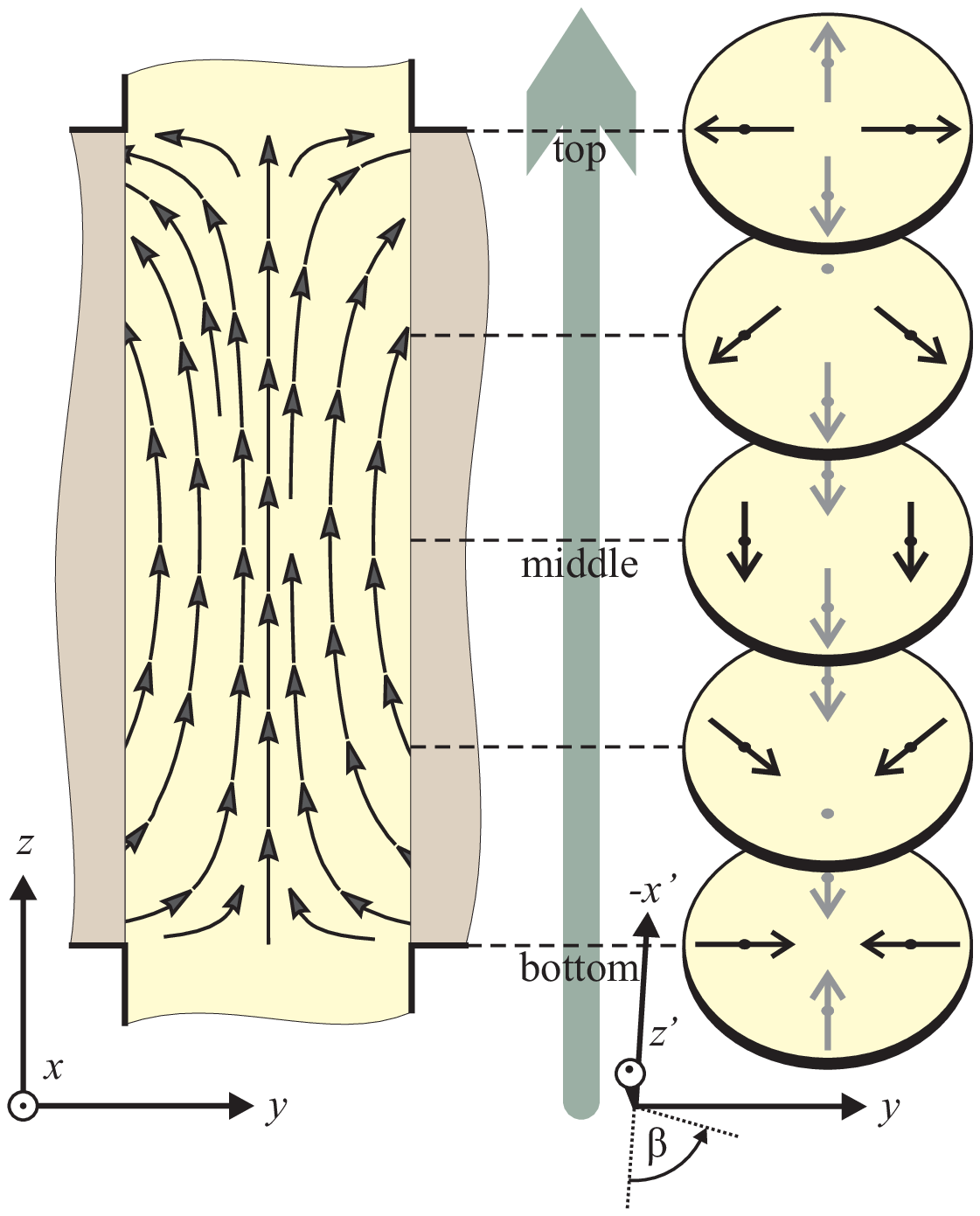}
  \end{center}
  \caption{A tilt of the C-field results in a position dependent rotation of the \rf field during the cavity passage. All fields are shown for $t=0$. The cross-sections on the right show the projections in  the $x'y$ plane defined by a C-field tilt towards $-x$. The arrows illustrate the local direction (as given by $\beta$) of the cavity field, but not its amplitude.
As described in the text, there is a clockwise rotation of the field direction for an upwards passage at $y<0$, and a counter-clockwise rotation at $y>0$.\newline
Note that the origin of the coordinate system is at the center-point of the cavity and the arrows shown just indicate axis orientation.}
\label{fig_crosssections}
\end{figure}

If the C-field is tilted by an angle $\alpha$ relative to the vertical $z$-axis, the \rf components that are orthogonal to the field depend not only on $H_x$ and $H_y$, but also on $H_z$. We choose the coordinate system such that the C-field tilt is towards the negative $x$-direction, as shown in \fref{fig_tilted_coordinates}. Additionally, we define a coordinate system $(x',y,z')$ that is tilted around the common $y$-axis by the tilt angle $\alpha$, such that the \trans{0}{-1} transition is now driven by the unchanged $H_y$ component and $H_{x'}= H_{x} \cos \alpha + H_{z} \sin \alpha$. Since $H_{z}$ is generally much stronger than $H_{x}$, it can have a significant effect even for small tilt angles. 

This is illustrated in \fref{fig_crosssections}
: In the cavity's center plane at $z=0$, the horizontal components $H_x$ and $H_y$ are zero. The field is then given by the projected $H_z$ component and therefore oriented towards $+x'$ everywhere. Near the top and bottom of the cavity the horizontal components dominate over the weak $H_z$ field and the radial symmetry is mostly preserved: The field points inwards at the bottom of the cavity and outwards at the top. The transitions between the top, center and bottom regions are smooth, which generally causes the direction of the linearly polarized field experienced by an atom to rotate during a cavity passage. For an upwards passage at $y < 0$, the rotation is clockwise, resulting in a positive shift of the effective microwave frequency driving the \trans{0}{-1} transition, as discussed previously. While the rotation in most cases does not occur at a constant angular velocity, this still breaks the symmetry of the spectrum around the resonant frequency and tends to lead to an enhancement of the lobe near $-\omega_\mathrm{rot}$. For a traversal at $y > 0$ the rotation occurs in a counter-clockwise direction instead, leading to an enhanced lobe near $+\omega_\mathrm{rot}$.

The reversal occurs without rotation only for a passage at $y=0$, where the projected $H_z$ component is parallel to the radial component, such that its only effect is to shift the point of zero field amplitude away from $z=0$.


\section{Analytical model}
\label{sec_analytical_model}

In the following, we will develop a model based on a first-order quantum mechanical treatment and the analytical closed-cavity fields. This shows a simple way to quantify the asymmetry of the \trans{0}{-1} spectra, which allows us to extract the center-of-mass position for the subset of atoms contributing to frequency measurements.


\subsection{Calculating transition probability spectra}

\label{sec_spectrum_for_0to-1}
\label{sec_complex_Rabi}

For low transition probabilities, the Bloch equations can be solved approximately by assuming a non-depleted initial state population (see \aref{sec_QM_derivation}). The resulting transition probability spectrum is then given by
\begin{equation}
\label{eq_P_from_FT}
P = \frac{c_t}{4} \abs{\func{\widetilde{\Omega} }{\Delom}}^2
\ ,
\end{equation}
where $\func{\widetilde{\Omega} }{\Delom}$ is the Fourier transform of the applied pulse given in terms of the complex Rabi frequency $\func{\Omega}{t} = \abs{\func{\Omega}{t}} e^{- i \, \varphi}$, where $\varphi$ describes the phase of the effective \rf field. The coefficient $c_t$ allows us to handle the variation of transition strengths~\cite{Vanier1989}: For simplicity, we will express all field amplitudes in terms of the Rabi frequency for the \trans{0}{0} transition, such that $c_{00} = 1$. For the \trans{0}{-1} transition driven by a linearly polarized \rf field orthogonal to the C-field, the transition strength coefficient is $c_{01}=5/16$~\footnote{The transition strength coefficient for \trans{-1}{0} is $c_{10}=3/16$.}, which already takes into account that only one of the circular polarization contributions has an effect.

In the presence of a C-field tilt $\alpha$ (see \fref{fig_tilted_coordinates}), the orthogonal pulse component takes the form
\begin{equation}
  \label{eq_pulse_01}
  \func{\Omega_\perp }{t} 
  = \func{\Omega_x}{t} \cos \alpha + \func{\Omega_y}{t} + \func{\Omega_z}{t} \sin \alpha
  \ .
\end{equation}
The projected components of $\Omega_x$ and $\Omega_z$ are oriented towards the $x'$-direction and their phase will serve as a reference. Relative to this, the linear polarization of the $\Omega_y$ component is oriented at an angle of $\beta = \pi / 2$. The resulting negative phase shift $\varphi=-\pi / 2$ for the \trans{0}{-1} transition then yields an extra factor of $i$ for the complex Rabi frequency $\Omega_y$.

For a single vertical passage, assumed to occur at constant velocity, the individual pulse components can now be expressed based on \eref{eq_amp_Hrho} and \eref{eq_amp_Hz} as
%
%
\begin{eqnarray}
\label{eq_Omega_x}
&\func{\Omega_x}{t}
= \ &\dfrac{\pi \  d \ b \ \zeta}{2 \, \tcav \cos \alpha} \,
   \dfrac{\pi^2}{2 \, \gamma \, d^2} \,
   \dfrac{x}{\rho} \,
   \func{J_1}{ \gamma \rho } \,
   \func{\sin}{ \dfrac{\pi \, t}{\tcav} }
   \func{\sqcap}{ \dfrac{t}{\tcav} }
\\
\label{eq_Omega_y}
&\func{\Omega_y}{t}
= i &\dfrac{\pi \  d \ b \ \zeta}{2 \, \tcav \cos \alpha} \,
   \dfrac{\pi^2}{2 \, \gamma \, d^2} \,
   \dfrac{y}{\rho} \,
   \func{J_1}{ \gamma \rho  } \,
   \func{\sin}{ \dfrac{\pi \, t}{\tcav} }
   \func{\sqcap}{ \dfrac{t}{\tcav} }
 \\
\label{eq_Omega_z}
&\func{\Omega_z}{t} 
= \ &\dfrac{\pi \  d \ b \ \zeta}{2 \, \tcav \cos \alpha} \,
   \dfrac{\pi}{2 \, d} \,
   \func{J_0}{ \gamma  \rho } \,
   \func{\cos}{ \dfrac{\pi \, t}{\tcav} } \,
   \func{\sqcap}{ \dfrac{t}{\tcav} } 
\ ,
\end{eqnarray}
%
%
where the rectangle function $\sqcap$ ensures that there is no effect of the \rf signal when the atom is outside the cavity. The common amplitude factor $\left( \pi \, d \, b \, \zeta \right) /\left( 2 \, \tcav \cos \alpha \right)$ is chosen such that for a resonant pulse with $b=1$, $\zeta=1$ and $\alpha=0$, an atom passing the cavity at $\rho=0$ experiences a $\pi/2$ pulse area for the \trans{0}{0} transition. The meaning of the amplitude factor $b$ and the cavity specific constant $\zeta$ is elucidated below. 

The Fourier-transforms of the pulse components can now be calculated as
\begin{eqnarray}
\label{eq_FT_Omega_x}
&\func{\widetilde{\Omega}_{x}}{\Delom}
=& \, i \, \dfrac{x}{\rho} \, \dfrac{b \, \zeta \, \pi^3}{8 \, d \, \gamma \, \cos \alpha} \, \func{J_1}{\gamma \rho } \Xfunc
\\
\label{eq_FT_Omega_y}
&\func{\widetilde{\Omega}_{y}}{\Delom}
=& - \dfrac{y}{\rho} \,  \dfrac{b \, \zeta \, \pi^3}{8 \, d \, \gamma \, \cos \alpha} \, \func{J_1}{\gamma \rho } \Xfunc
\\
\label{eq_FT_Omega_z}
&\func{\widetilde{\Omega}_z}{\Delom}
=& \dfrac{b \, \zeta \, \pi^2}{8 \, \cos \alpha} \, \func{J_0}{\gamma \rho } \Yfunc
\ ,
\end{eqnarray}
where
\begin{equation}
\label{eq_definition_Xfunc}
\Xfunc 
:= \funcSq{\sinc}{\frac{\tcav}{2} \left(\Delom + \frac{\pi}{\tcav} \right)}
    - \funcSq{\sinc}{\frac{\tcav}{2} \left(\Delom - \frac{\pi}{\tcav} \right)} 
\end{equation}
and
\begin{equation}
\label{eq_definition_Yfunc}
\Yfunc
:= \funcSq{\sinc}{\frac{\tcav}{2} \left(\Delom + \frac{\pi}{\tcav} \right)}
    + \funcSq{\sinc}{\frac{\tcav}{2} \left(\Delom - \frac{\pi}{\tcav} \right)}
\end{equation}
describe the dependence on the detuning $\Delom$.
%
%
Due to the linearity of the Fourier transform, the \trans{0}{-1} transition probability given by \eref{eq_P_from_FT} now depends on the linear combination of the components according to \eref{eq_pulse_01} as 
\begin{eqnarray}
\label{eq_linear_combination_with_tilt}
\func{P_{01}}{\Delom}  
= \frac{c_{01}}{4} \left(b \, \zeta \right)^2
  \Big|&&
    \left(\frac{i \, x}{\rho} - \frac{y}{\rho \cos \alpha} \right)
    \frac{\pi^3}{8 \, d \, \gamma} \, \func{J_1}{\gamma \rho } \Xfunc
\nonumber \\    
    &&+ \frac{\pi^2}{8} \, \func{J_0}{\gamma \rho } \Yfunc \, \tan \alpha 
  \Big|^2
\ . 
\end{eqnarray}
For $\alpha = 0$ the expression reverts to the expected radial symmetry.

We now explicate the role of the factors $b$ and $\zeta$: These describe the actual \rf amplitude in relation to the value used in normal fountain operation, where it is optimized to maximize the contrast of the Ramsey fringes of the \trans{0}{0} transition. This transition is excited by the pulse component parallel to the C-field
\begin{equation}
\label{eq_omega_parallel}
  \func{\Omega_\parallel}{t} = \func{\Omega_z}{t} \cos \alpha - \func{\Omega_x}{t} \sin \alpha
  \ .
\end{equation}
As shown in \aref{sec_QM_derivation}, the resonant tipping angle after a passage through the assumed \rf field distribution in the cavity is then given by
\begin{eqnarray}
  \Theta
  & = \abs{ \func{\widetilde{\Omega}_\parallel }{0} }
    = \abs{ \func{\widetilde{\Omega}_z }{0} } \cos \alpha 
\label{eq_tipping_angle_gamma_rho}
  & = b \, \zeta \tfrac{\pi}{2} \, \func{J_0}{\gamma \rho }
  \ ,
\end{eqnarray}
where no $\widetilde{\Omega}_x$ term appears since $\Xfunc = 0$ for $\Delom = 0$. Averaged over the entire atomic cloud, this gives a transition probability after the first cavity passage of
%
\begin{eqnarray}
  \widehat{P}_{00}
  \label{eq_P00_avg}
  &= \frac{1}{2} \, \int_0^{r_a}       
    \left(
      1 - \func{\cos}{ b \, \zeta \frac{\pi}{2} \, \func{J_0}{\gamma \rho } \, }
    \, \right)
    \func{W_r}{\rho}
    d\rho
    \ ,
\end{eqnarray}
where the normalized distribution $\func{W_r}{\rho}$ describes the probability of finding any given atom at a position with the radial coordinate $\rho$.

\begin{figure}[t!b]
  \begin{center}
    \includegraphics[width=86 mm]{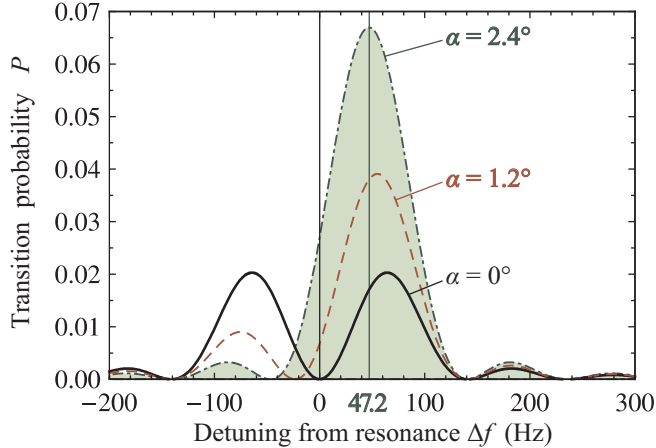}
  \end{center}
  \caption{
Variation of the \trans{0}{-1} spectrum with C-field tilt angle $\alpha$ for a single atom traversing the cavity at a fixed horizontal position $\vec{r}$, calculated using \eref{eq_linear_combination_with_tilt}.
{\em Heavy black curve:} At $\alpha=0^\circ$, the spectrum is symmetric around resonance, with two lobes of equal amplitude near $\pm \omega_\mathrm{rot} = 2 \pi \times \un{47.2}{Hz}$.
{\em Green shaded, dot-dashed curve:} The horizontal position was chosen such that at $\alpha=2.4^\circ$ the field direction undergoes a rotation of constant angular velocity, resulting in a spectrum that is again symmetric, but with a single central lobe shifted by $\omega_\mathrm{rot}$.
{\em Red, dashed curve:} For an intermediate value of $\alpha=1.2^\circ$, the lobes of the spectrum remain at similar positions as without tilt, but the transition probabilities are enhanced around $\omega_\mathrm{rot}$.}  
\label{fig_spectra_shifted_and_real}
\end{figure}

Assuming no correlation between the atomic positions during the first and second passage, a local maximum of the fringe contrast as a function of \rf amplitude is achieved whenever $\widehat{P}_{00} = 0.5$. In analogy with \cite{Li2010}, the cavity specific constant $\zeta$ is now chosen such that the first of these maxima occurs at $b=1$ for a uniform density distribution (where $\func{W_r}{\rho}=2 \rho \, / \,r_a^2$). For the cavity used in CSF2, this yields $\zeta=1.080$.

The density in a realistic atomic distribution is generally higher in the stronger field near the cavity axis, and a detailed analysis for CSF2~\cite{Weyers2012} yields a value of $b=0.97$ for the first maximum of the contrast. The experiments presented throughout the paper were performed at a nominal $5 \pi / 2$ pulse area ($b_\mathrm{nom}=5$), for which we assume a five times larger amplitude, as described by $b=4.85$.

Note that the \rf amplitude is adjusted for maximum contrast while the C-field tilt is already present. The necessary compensation for the reduced field component parallel to the C-field that can be seen in \eref{eq_omega_parallel} then causes the field amplitudes given by \eref{eq_Omega_x}--\eref{eq_Omega_z} to increase with the value of $\alpha$ even if $b$ is constant.

The \trans{0}{-1} transition probability spectrum for any given microwave amplitude, C-field tilt, and horizontal position of cavity passage can now be found from \eref{eq_linear_combination_with_tilt} with an accuracy that is limited by the non-depletion approximation and the assumed analytical field distribution. Examples are shown in \fref{fig_spectra_shifted_and_real}.


\subsection{Estimating the angle of C-field tilt}
\label{sec_tilt_angle}

A point of particular interest in the spectrum of the \trans{0}{-1} transition is the resonant transition probability $P_{01}^\mathrm{res} = \func{P_{01}}{0}$. Since $\Xfunc=0$ for $\Delom=0$, this depends only on the last term of \eref{eq_linear_combination_with_tilt} as
\begin{equation}
\label{eq_P01_res}
P_{01}^\mathrm{res}  
= \left( c_{01} / 4 \right)
  \left(
    b \, \zeta \,
    \tfrac{\pi}{2} \, \func{J_0}{\gamma \rho } \, \tan \alpha 
  \right)^2
\ .
\end{equation}
%
The non-depletion approximation is quite accurate here since the resonant transition probability for the \trans{0}{-1} transition is generally very low. 

As shown in \aref{sec_tilt_derivation}, the C-field tilt angle $\alpha$ can then be calculated from the resonant transition probability $\widehat{P}_{01}^\mathrm{res}$ averaged over the entire cloud as
%
%
%
%
\begin{equation}
\label{eq_tilt_from_Pres}
\alpha
= \func{\arctan}{
  \frac{4}{\pi \, b_\mathrm{nom}}
  \sqrt{ \widehat{P}_{01}^\mathrm{res} / c_{01}} \,
}
\ ,
\end{equation}
even if the exact details of the density distribution in the cavity are not known.

When the spectrum is symmetric, $\widehat{P}_{01}^\mathrm{res}$ is simply the transition probability at the central minimum and therefore easy to determine. This is not true for the asymmetric case, where the minimum appears shifted away from resonance (see \fref{fig_spectra_shifted_and_real} and \fref{fig_sample_spectra}). While it is possible to find the proper resonant frequency from a non-linear fit of the spectrum, it is preferable to obtain it without introducing a new free parameter by measuring the resonant frequency of the \trans{-1}{-1} transition and using the Breit-Rabi formula~\cite{Breit1931}. A series expansion yields the relevant terms
\begin{equation}
  \label{eq_linearized_Breit_Rabi}
  \Delom_\mathrm{[0\, to\, -1]}
  = \left(\tfrac{1}{2}+4\, g_I / g_J\right) \Delom_\mathrm{[-1\, to\, -1]} \ ,
\end{equation}
where $g_I$ and $g_J$ are the nuclear and electronic $g$-factors, respectively, and the resonant frequencies $\Delom_\mathrm{[0\,to\,-1]}$ and $\Delom_\mathrm{[-1\,to\,-1]}$ are given relative to the \trans{0}{0} clock transition. Note that the magnetic field in the cavity can differ noticeably from the value averaged over the entire atomic trajectory that is measured in atomic fountains to correct for the quadratic Zeeman shift.

 
\subsection{Extracting position data from measurements}
\label{sec_measuring_position}

Spectra calculated using \eref{eq_linear_combination_with_tilt} show that the asymmetry mostly depends on $y$. In this section we develop this into a method to extract the center-of-mass position along the $y$-axis of the atomic cloud in the cavity from a measured spectrum.

Summing real-valued and imaginary parts of \eref{eq_linear_combination_with_tilt} in quadrature yields
%
%
%
%
\begin{eqnarray}
  \func{P_{01}}{\Delom} &&= \frac{c_{01}}{4} \left(b \, \zeta \right)^2
\nonumber \\
    \times \Bigg[
      &&\left(
        \frac{x^2}{ \rho^2} + \frac{y^2}{ \rho^2 \cos^2 \alpha}
      \right)
      \left(
        \frac{\pi^3}{8\, d \, \gamma} \,
        \func{J_1}{ \gamma \rho }
      \right)^2
      \Xfunc^2
 \nonumber \\
    &&+ \left(
        \frac{\pi^2}{8} \func{J_0}{ \gamma \rho }
        \tan \alpha
      \right)^2
      \Yfunc^2
 \nonumber \\
   &&- 2 \left(    
        \frac{y}{ \rho } \frac{\sin \alpha}{\cos^2 \alpha} 
        \frac{\pi^5}{64 \, d \, \gamma} 
        \func{J_1}{ \gamma \rho }      
        \func{J_0}{ \gamma \rho }        
      \right)
      \Xfunc \, \Yfunc
    \Bigg]
    .
\end{eqnarray}
The last term with its dependency on $y$ is the most interesting and can be isolated by exploiting the different symmetries of the terms: Since $\func{\Xfunc}{-\Delom} = -\func{\Xfunc}{\Delom}$ and $\func{\Yfunc}{-\Delom} = \func{\Yfunc}{\Delom}$, the terms containing $\Xfunc^2$ and $\Yfunc^2$ are symmetric in $\Delom$, while the final term depends on $\left(\Xfunc \, \Yfunc\right)$ and is antisymmetric. The unwanted terms can be removed by defining the asymmetry of the transition probability spectrum as
%
\begin{eqnarray}
\label{eq_dP_definition}
\func{\dP}{\Delom}
&&= \tfrac{1}{2} \left( \func{P_{01}}{-\Delom} - \func{P_{01}}{\Delom} \right)
\\
&&= y \left(b \, \zeta \right)^2 
    \frac{\sin \alpha}{\cos^2 \alpha}
    \frac{c_{01} \, \pi^5}{256 \, d}
    \left\{
      \frac{
        2 \, \func{J_1}{ \gamma \rho }      
        \func{J_0}{ \gamma \rho }
      }{ \gamma \rho }
    \right\}
    \Xfunc \, \Yfunc 
    \ .
\nonumber \\
\label{eq_dP_equation}
\end{eqnarray}
The term in curly brackets approximates to $1$ when the radial coordinate $\rho$ is small, showing the asymmetry $\dP$ to be near-linear in the $y$-coordinate of the cavity passage.

Any measured asymmetry value $\dPhat$ obtained in an atomic fountain is an average of many individual atomic spectra in the same way as in \eref{eq_P00_avg}, except that the assumption of radial symmetry is no longer valid here:
\begin{equation}
\label{eq_dPhat_equation}
  \dPhat
  = \mathop{\int \!\!\! \int}_{\mathrm{A_{ac}}} \func{W}{x,y} \func{\dP}{x,y} dx \, dy
  \ .
\end{equation}
The normalized density distribution $\func{W}{x,y}$ is defined to be zero outside the accessible area $A_\mathrm{ac}$, given by the radius $r_a$ of the cutoff tubes.

If $\dP$ were entirely linear in $y$ and independent of $x$, this could be rewritten as the product of a constant slope factor $s$ and the center-of-mass coordinate $\widehat{y}$ for the ensemble of detected atoms:
\begin{equation}
\label{eq_yhat_equation}
  \dPhat   
  = s \times \widehat{y}
  \quad \mathrm{with \ }
  \widehat{y} = \mathop{\int \!\!\! \int}_{\mathrm{A_{ac}}} \func{W}{x,y} \, y \ dx \, dy  
  \ .
\end{equation}
The position $\widehat{y}$ could then easily be determined by taking the asymmetry of a measured single-passage spectrum and dividing it by the value calculated for $s$.

This approach does in fact provide useful results as experimentally confirmed in \sref{sec_experimental_test} and discussed further in \sref{sec_centering_uncertainty}: The deviation of $\dP$ from linearity is quite small, and its local variation is partially averaged out since the atomic cloud generally has a smooth density distribution throughout the cavity, such that the variation of $\dPhat$ with $\widehat{y}$ can be described through an effective slope value $s_\mathrm{eff}$.

Measuring $\widehat{y}$ is then possible with the following procedure: A C-field tilt along the $x$-axis is applied. The fountain is configured such that a microwave pulse is applied only during a single cavity passage and ideally the microwave amplitude is reoptimized accordingly. The transition probability $\widehat{P}_{01}$ is measured on resonance and at two symmetric detunings $\Delom=\pm \Delom_m$. The resonant transition probability $\widehat{P}^\mathrm{res}_{01}$ yields a value for the C-field tilt $\alpha$~\eref{eq_tilt_from_Pres}, which is used to find the slope factor $s_\mathrm{eff}$. 
Finally, the center-of-mass $y$-position is extracted from the asymmetry as $\widehat{y} = \dPhat / s_\mathrm{eff}$. The coordinate $\widehat{x}$ can be obtained in the same way by applying a C-field tilt along the $y$-axis.


\section{Experimental test}
\label{sec_experimental_test_main}

\begin{figure}[bt]
  \begin{center}
    \includegraphics[width=86 mm]{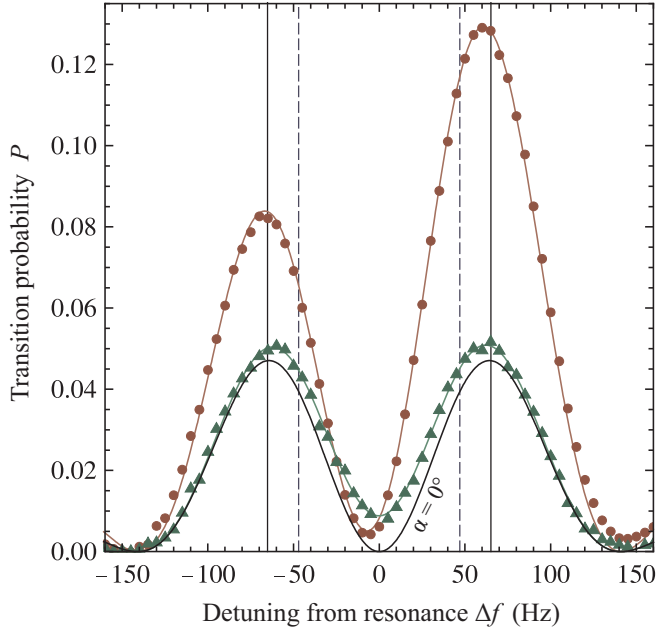}
  \end{center}
  \caption{
Comparison of experimental (symbols) and calculated single-passage spectra (lines).
Experimental data was taken with the initial cloud deliberately placed off-center (red circles) and with the molasses beam balance adjusted for improved centering (green triangles). This data was then fitted with~\eref{eq_linear_combination_with_tilt} by varying the horizontal position of passage $\vec{r}$ for a C-field tilt of $2.4^\circ$ extracted from the resonant transition probability (red, green lines). To show the effect of the C-field tilt, the spectrum for centered adjustment was recalculated assuming $\alpha=0$ (black line).
Two choices of the measurement detuning (see \sref{sec_measurement_detuning}) have been marked. Dashed vertical lines indicate $\omega_m = \pm \omega_\mathrm{rot}$ (maximum sensitivity) and solid lines indicate $\omega_m = \pm 4.3 / \tcav$ (suppressed effect of frequency uncertainty).
}  
\label{fig_sample_spectra}
\end{figure}

\subsection{Shape of observed spectra}
\label{sec_experimental_spectra}

Even without knowledge of the exact value of $s_\mathrm{eff}$, the measured spectral asymmetry can be used to adjust the position of the launched cloud in the cavity. We used this method for preliminary centering of the cloud while loading atoms from a slow caesium beam. Since there is no centering force in the optical molasses employed in CSF2, the unidirectional loading from the beam tends to create an asymmetric, off-center cloud. By investigating the asymmetry of the spectra obtained with a microwave signal applied during either the upwards or the downwards passage, both the initial position of the cloud and the launch direction can be adjusted in this way.

\Fref{fig_sample_spectra} shows spectra taken during the centering process, one for a far off-center position and one after adjustment. The microwave pulse was applied during the upwards passage, with an amplitude that corresponds to $b_\mathrm{nom}=5$. The C-field tilt was calculated as $\alpha=2.4^\circ$ from the resonant transition probability. Note that the C-field tilt inside the cavity of CSF2 so far cannot be deliberately controlled and simply results from the superposition of stray magnetic fields with the homogeneous, vertical field produced by the C-field coil.

To demonstrate the adequacy of the analytical model, the figure also shows a non-linear fit of each spectrum with \eref{eq_linear_combination_with_tilt}, which describes the spectrum for a single atom. The best agreement is achieved for a cavity passage at a horizontal position with $y^\mathrm{off}=\un{1.9}{mm}$ in the off-center case and $y^\mathrm{cent}=\un{0.0}{mm}$ after centering.


\label{sec_measurement_detuning}

The highest sensitivity for a position measurement is obtained by measuring at a detuning $\Delom_m = \omega_\mathrm{rot}$, where $\Xfunc \, \Yfunc = -1$, maximizing the value of $\dP$.
However, as shown by the dashed vertical lines in \fref{fig_sample_spectra}, the transition probability is then measured in a region of the spectrum with a considerable slope. This translates any error in the determined resonant frequency into an error of $\dPhat$ that can be large: In the off-center example shown in the figure, a realistic frequency uncertainty of $\sigma_\mathrm{res}=\un{2\pi \times 1}{Hz}$ causes an uncertainty contribution of $\sigma_{\!\dP[\omega]} = 1.5 \times 10^{-3}$, which is considerably larger than the statistical measurement uncertainty.

A good choice for an optimized measurement detuning is $\Delom_m=4.3 / \tcav$, such that the measurements are effectively taken at the maxima of $\Xfunc$ (see \eref{eq_definition_Xfunc}), which provides the largest contributions to the transition probability spectrum. Using this for the sample measurement reduces the uncertainty contribution to an insignificant value, 
at the price of reducing the position sensitivity by $13\%$.

\subsection{Variation of $\dP$ with launch direction}
\label{sec_experimental_test}

To further confirm the calculations, a series of experiments was performed. One of the design features of the fountain CSF2 is a goniometric stage for the entire molasses zone that makes it possible to vary the launch direction in a highly reproducible fashion without tilting the entire fountain~\cite{Gerginov2010}.

Measurements of the detected atom number, the resonant transition probability $\widehat{P}^\mathrm{res}_{01}$ and the asymmetry $\dPhat$ were taken for a range of launch angles up to $\un{3}{mrad}$, covering both the position sensitive $y$-direction and the insensitive $x$-direction. 
Since it is unknown which alignment results in a vertical launch, the tilt of the launch direction is given by the angles $\delta_x$ and $\delta_y$ relative to the direction that maximizes the returned atom number. This optimized tilt could be determined to within an uncertainty of $\sigma_\delta=\un{0.05}{mrad}$ by fitting the atom number detected as a function of the launch tilt in $x$- and $y$-direction with a 2D Gaussian distribution for a suitable subset of measurements. The measurement detuning for the determination of $\dPhat$ was chosen as $\Delom_m=2\pi \times \un{65.2}{Hz}$, based on the optimized value from~\sref{sec_measurement_detuning}. All measurements were performed twice, with the microwave pulse applied either during the upwards passage or during the downwards passage, and the results given are averaged over 100 successive launches.

After correcting for crosstalk in the detection system, the measured resonant transition probabilities $\widehat{P}^\mathrm{res}_{01}$ for the upwards and the downwards passage at each launch tilt all agree with the average value of $\widehat{P}^\mathrm{res}_\mathrm{avg} = 2.86 \times 10^{-3}$ to within a measurement uncertainty of $\sigma_{\!P} = 5 \times 10^{-4}$. For a nominal pulse area described by $b_\mathrm{nom}=5$ and using~\eref{eq_tilt_from_Pres}, this result corresponds to a tilt angle of $\alpha=1.4^\circ$. The difference to the tilt angle of $2.4^\circ$ found in experiments conducted several months earlier (such as the one in \sref{sec_experimental_spectra}) indicates a change in the stray magnetic fields that is likely due to work on the attached caesium beam apparatus over the intervening period.

Looking at the spectral asymmetries $\func{\dPhat}{\delta_x,\delta_y}$ for either the upwards or the downwards cavity passage, a three-dimensional plot of the data points obtained shows them to lie within a plane. As the tilt of the C-field used in the measurements was originally unintended and of unknown direction, the orientation of the plane fitted to the data for the downwards passage was used to align the coordinate system such that a change of the launch tilt $\delta_y$ (along the $y$-axis) results in the largest change of $\dPhat$.

\begin{figure*}[bt]
  \begin{center}
    \includegraphics[width=170 mm]{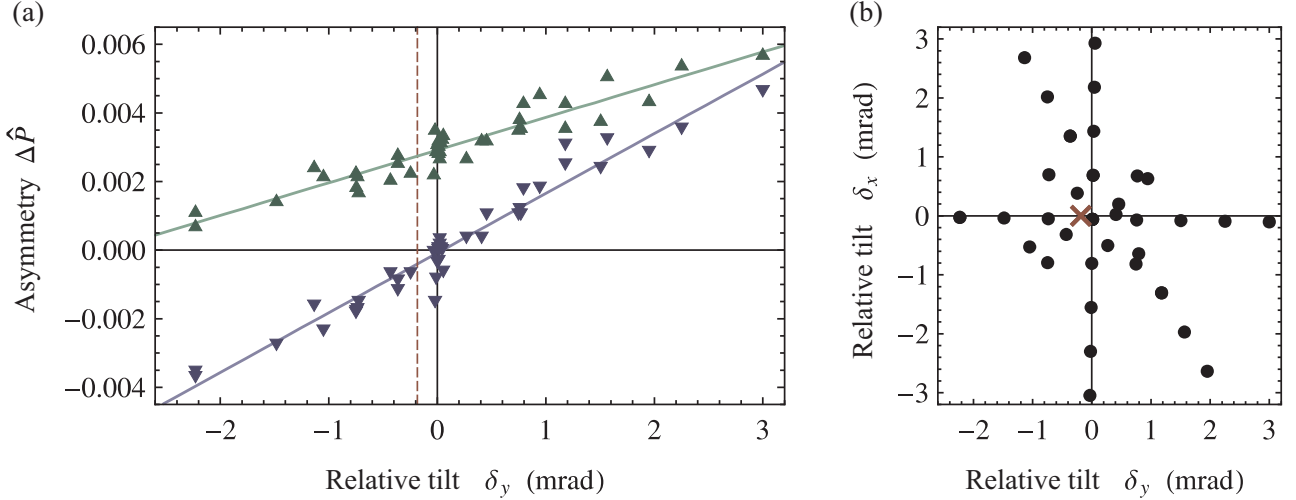}    
  \end{center}
  \caption{
(a) Measured asymmetry values $\dPhat$ over launch tilt $\delta_y$ in $y$-direction together with linear fits. Green upward triangles mark data taken for the upwards passage, blue downward triangles for the downwards passage. The red dashed line marks the tilt setting for a vertical launch direction obtained from the simulation. (b) Combinations of launch tilts $\delta_x$ and $\delta_y$ where data was taken. The red x marks the direction of vertical launch.
}
\label{fig_deltaP_measurements}
\end{figure*}

\Fref{fig_deltaP_measurements}(a) can be regarded as a cross-section of the three-dimensional plot: It shows $\dPhat$ for all investigated launch angles plotted over the relative launch tilt $\delta_y$. The deviations from the linear fit are then independent of $\delta_x$ and consistent with a purely statistical variation with standard deviations of $\sigma_{\dP}^\uparrow = 4.3\times 10^{-4}$ and  $\sigma_{\dP}^\downarrow = 3.6\times 10^{-4}$ for the upwards and downwards passages. The slight increase over the value expected from simple uncertainty propagation, $\sigma_{\dP} = \sigma_{P} / \sqrt{2} \approx 3.5 \times 10^{-4}$, is attributed to the decrease in detected atom number at large launch tilts and the resulting reduction in signal-to-noise ratio.

Using the effective slope values $s_\mathrm{eff}^\uparrow=\un{-6.35}{m^{-1}}$ and $s_\mathrm{eff}^\downarrow=\un{6.26}{m^{-1}}$ (obtained in \sref{sec_numerical_slope_value}) now yields a linear fit for the center-of-mass positions $\widehat{y}$ as a function of $\delta_y$:
%
\begin{eqnarray}
\label{eq_lin_shift_up}
\func{ \widehat{y}^\uparrow }{\delta_y} 
  &= \un{-0.47}{mm} - \left( \un{0.14}{mm/mrad} \right) \delta_y \\
\label{eq_lin_shift_dn}
\func{\widehat{y}^\downarrow }{\delta_y} 
  &= \un{-0.01}{mm} + \left( \un{0.28}{mm/mrad} \right) \delta_y
\end{eqnarray}
for the upwards and downwards cavity passage, respectively.

The fact that $\func{ \widehat{y}^\uparrow }{\delta_y} < 0$ for all tested launch tilts
indicates that the cloud in the molasses zone is positioned off-center even before launch. Due to the lack of a centering force, this can easily happen if the cooling beams are not perfectly aligned. Adjusting the launch tilt then mostly affects the cloud position during the downwards passage $\func{\widehat{y}^\downarrow }{\delta_y}$, which the measurements confirm to be well-centered at $\delta_y=0$, where the maximum returned atom number is detected.

Interestingly, the cloud position on the way up shifts against the launch tilt due to a position dependent selection process: For a launch tilt $\delta_y>0$, atoms with an initial position $y > 0$ are more likely to drift out of the accessible area before the second cavity passage and are therefore less likely to be detected. Conversely, atoms placed at $y < 0$ now tend to move towards the fountain axis which increases their chance of detection. For the ensemble of atoms that eventually get detected, the center-of-mass position during the upwards cavity passage then appears shifted towards $-y$, against the launch tilt (see also section 6 of~\cite{Li2010}). This shift is not compensated by the additional horizontal motion between launch and first cavity passage unless the cloud is very small.

The same selection process also causes a general suppression of the cloud shift with tilt: For $\delta_y>0$, atoms with an opposite thermal velocity component $v_y^\mathrm{th}<0$ are more likely to reach the detection zone than those with $v_y^\mathrm{th}>0$ and therefore contribute more to the detected ensemble. This is the reason why the observed shift of the cloud between passages of $\un{0.42}{mm}$ per mrad launch tilt is much smaller than the value of $\un{2.5}{mm}$ per mrad expected for a simple parabolic trajectory.


\subsection{Numerical simulation}
\label{sec_trajectory_simulation}
\label{sec_numerical_slope_value}

To confirm the effect of these selection mechanisms on the observed center-of-mass positions, a numerical trajectory simulation was developed. It is based on Monte-Carlo methods and generates atomic trajectories starting from a three-dimensional cloud with a Gaussian distribution of initial positions and velocities. Trajectories are rejected if they intersect any of the known restrictions, most importantly the cutoff tubes of state selection and Ramsey cavities, or if they miss the detection zone.

For each remaining trajectory, the horizontal position of the upwards or downwards cavity passages is determined, such that $\dPhat$ for the upwards and downwards passage can be obtained from the average transition probabilities at the detunings $\pm \Delom_m$. To eliminate the error introduced by the non-depletion approximation, the transition probabilities are found by numerically solving the differential equations \eref{eq_DE1} and \eref{eq_DE2} instead of using \eref{eq_linear_combination_with_tilt}.

The best agreement with the experimental data for $\dPhat$ and the detected atom number, both as a function of launch tilt, is achieved for an initial cloud with a horizontal density distribution described by $\sigma_r=\un{3.0}{mm}$, placed at a horizontal position offset of $y_0=\un{-1.1}{mm}$. The horizontal velocity distribution is given by $\sigma_v=\un{7.9}{mm/s}$, corresponding to a temperature of $T_\mathrm{cloud}=\un{1.0}{\mu K}$.

Even at this very low cloud spread and for optimized launch direction, less than $15\%$ of the launched atoms are detected at the end of the cycle. Any change in the launch tilt then strongly affects which fraction of the cloud is most likely to be detected, confirming the counter-intuitive shift of the cloud on the upwards cavity passage and the reduced shift on the downwards passage.

After optimization, the trajectory simulation also provides values for the effective slope value $s_\mathrm{eff}$ (compare \sref{sec_measuring_position}): For a range of launch tilts varying in both $x$- and $y$-direction, we calculate the center-of-mass position $\widehat{y}$ of the simulated atomic ensemble in addition to $\dPhat$. By plotting $\dPhat$ over $\widehat{y}$, the effective values $s_\mathrm{eff}^{\,\uparrow}=\un{-6.35}{m^{-1}}$ and $s_\mathrm{eff}^{\,\downarrow}=\un{6.26}{m^{-1}}$ for the upwards and downwards passage are then obtained from linear fits. The sign reversal for $s_\mathrm{eff}^{\,\downarrow}$ results from the vertical antisymmetry of $H_\rho$.


\subsection{Assumed density distribution}

\label{sec_density_distribution}

The Monte-Carlo nature of the trajectory simulation complicates a systematic evaluation of the position uncertainty arising from the distribution of atoms in the cavity. Because of this, we introduce a simplified model to describe the density distribution inside the cavity during the upwards or downwards passage, for those atoms that are detected at the end of the cycle and are therefore relevant for frequency measurements in the fountain.

The trajectory simulation shows that due to clipping at the cutoff tubes and the limits of the detection zone, the density of these relevant atoms falls off sharply in the region near the edge of the accessible area, so that we can describe the relevant atomic density by the approximate distribution
\begin{eqnarray}
  &&\func{W_a}{x,y} 
  = \left\{ 
    \begin{array}{cl}
      a_n \exp -\dfrac{\left( x - x_c \right)^2 + \left( y - y_c \right)^2}{2 \sigma^2}
      & : \rho \le r_\mathrm{lim} 
      \\
      0 
      & : \mathrm{else}
      \end{array}  
  \right.
  \nonumber \\
  &&\quad \mathrm{with} \quad \rho = \sqrt{x^2 + y^2}
  \ ,
\end{eqnarray}
which is constrained to a radius $r_\mathrm{lim}$ around the cavity axis. The density fall-off within this radius (which is smaller than the cutoff tube radius $r_a$) is described by $\sigma$, while $x_c$ and $y_c$ provide a way to vary the center-of-mass position of the cloud. The normalization factor $a_n$ is calculated numerically. 

Values for $\dPhat$ and $\widehat{y}$ and  can now be found from \eref{eq_dPhat_equation} and \eref{eq_yhat_equation}. In the former, it is again preferable to calculate the local values of $\dP$ from a numerical solution of the differential equations. Since both $\widehat{y}$ and $\dPhat$ are zero for a centered cloud (where $x_c = y_c = 0$), the effective slope value for an exemplary deflection of the cloud is
\begin{equation}
\label{eq_effective_slope_factor}
  s_\mathrm{opt} = \frac{ \func{\dPhat}{x_c,y_c} }{ \func{\widehat{y}}{x_c,y_c} }
  \ .
\end{equation}
The exact values chosen for $x_c$ and $y_c$ are not critical, as the value obtained for $s_\mathrm{opt}$ changes by less than $0.5\%$ for center-of-mass positions within $\un{1}{mm}$ of the origin. In the following, we will assume $s_\mathrm{opt}$ to be calculated for a near-zero deflection.

The approximate distribution $W_a$ reproduces the slope values obtained directly from the trajectory simulation for a common limiting radius $r_\mathrm{lim}=\un{4.0}{mm}$, if ``unclipped'' cloud sizes of $\sigma^\uparrow = \un{3.3}{mm}$ and $\sigma^\downarrow = \un{7.3}{mm}$ are chosen for the upwards and downwards passage. These cloud sizes are in good agreement with the expected thermal expansion of the launched cloud.


\section{Uncertainty estimate}
\label{sec_errors}

Investigating the dependence of the spectral asymmetry on the cloud position was made possible by an uncontrolled and originally unintended tilt of the C-field in CSF2. This does not provide position information along the currently insensitive $x$-axis, and the accuracy of the position measurements along the $y$-axis is also not optimal. 

For the discussion of the achievable uncertainties, we will therefore assume a hypothetical version of CSF2 in which a set of coils provides an intentional tilt of the C-field that can be changed in both amount and direction. This allows operation at optimized parameters, which will be set as follows throughout the section:
Choosing an even value for $b_\mathrm{nom}$ makes it possible to adjust the \rf amplitude for maximum \trans{0}{0} transition probability after a single cavity passage, as the tipping angle is then a multiple of $\pi$. Compared to the optimization after both cavity passages that otherwise needs to be used, this agrees more closely with the assumptions used in the derivation of \eref{eq_tilt_from_Pres} and improves the accuracy of the determined C-field tilt $\alpha$. It is preferable to reach an even $b_\mathrm{nom}$ through a reduction of the nominal pulse area to $b_\mathrm{nom}=4$ (with $b\approx3.9$), since this also reduces the overall transition probability, which improves the linearity of $\dP$ in $y$. An increase of the C-field tilt to a value close to $\alpha = 3^\circ$ more than compensates the reduction of the slope value due to the reduced \rf amplitude to give $s_\mathrm{opt} \approx \un{8.9}{m^{-1}}$.


\subsection{Residual position uncertainty after centering}
\label{sec_centering_uncertainty}

\begin{figure}[t!bh]
  \begin{center}        
    \includegraphics[width=86 mm]{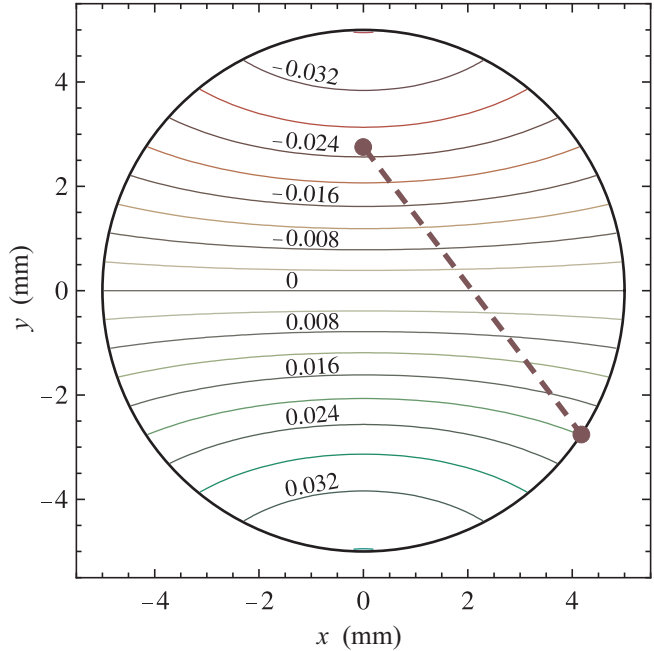}    
   \end{center}
  \caption{
Variation of the spectral asymmetry with the horizontal position of cavity passage given by $x$ and $y$. Values of $\func{\dP}{x,y}$ are given by the contour labels for optimized parameters $\alpha=3.0^\circ$, $b=3.9$ and $\Delom_m=2 \pi \times \un{65.2}{Hz}$. The variation of $\dP$ with $x$ can cause an error in the position determination from $\dPhat$ that is at its maximum for the situation illustrated by the marked points (see text).
}
\label{fig_contourplot}
\end{figure}

Ideally, the asymmetry measurements are used to center the atomic cloud in the cavity during both the upwards and downwards passage, and in both $x$- and $y$-directions. When $\dPhat$ is zero, the uncertainty of the calculated slope factor becomes insignificant. If the observed spectra resulted from a single atom, the residual uncertainty would then be $\sigma_{y}^\mathrm{sing} = \abs{s^{-1}_\mathrm{opt} \  \sigma_{\!\dP} } = \un{0.04}{mm}$.

Reducing $\sigma_{\!\dP}$ through increased averaging is only possible to the point where systematic effects such as the nonlinearity of the detection system or the transition probability background from the off-resonant excitation of the \trans{0}{0} transition become limiting. For the parameters used here, the latter was estimated to cause an error in the measured $\dPhat$ value on the order of $2\times 10^{-4}$, based on numerical calculations. 
At a similar level, the uncertainty resulting from a possible error in the assumed resonance frequency of the \trans{0}{-1} transition discussed in \sref{sec_measurement_detuning} might also need to be included in $\sigma_{y}^\mathrm{sing}$.

For an extended distribution of atoms there is another, larger contribution to the uncertainty: A value of $\dPhat=0$ is the result of positive and negative $\dP$ contributions from individual atoms. Since $\dP$ is not entirely linear in $y$ and also has a dependence on $x$, a density distribution with a center-of-mass position of $\widehat{y}=0$ does not necessarily yield $\dPhat=0$ and vice versa.

\Fref{fig_contourplot} illustrates this: The contour plot shows the variation of $\dP$ with the horizontal position of cavity passage for a single atom. Clearly, the same value of $\dP$ can occur for a range of $y$-positions if $x$ is varied. To quantify the resulting position uncertainty, we investigate a minimalistic distribution of just two atoms placed such that $\widehat{y}=0$ and therefore $y_1 = -y_2$. The largest deviation of $\dPhat$ from zero occurs when the atoms are placed as marked in the figure, one at $x_1=0$ and the other at the edge of the cutoff tube. The resulting asymmetry of $\dPhat=-2.7\times 10^{-3}$ corresponds to an error of the extracted center-of-mass position of $\Delta \widehat{y}=\un{0.31}{mm}$.

This two-position distribution is the worst-case arrangement and allowing other distributions with more atoms will not increase the resulting error. Fortunately the atomic density distribution normally is not concentrated in localized regions, but spread smoothly across the entire accessible area. Simply requiring a constant density distribution along the connecting line between the worst-case positions, as indicated in the figure, reduces the maximum position error to $\Delta \widehat{y}=\un{0.11}{mm}$, which we will use as the distribution based contribution to the uncertainty of the center-of-mass position after centering, $\sigma_{y}^\mathrm{dist}$.
%
Together with $\sigma_y^\mathrm{sing}$, this yields a combined centering uncertainty of $\un{0.12}{mm}$ for a measurement at optimized parameters. 
%
%


\subsection{Position uncertainty for an uncentered cloud}
\label{sec_slope_errors}

We now investigate how critical the assumed values for $\sigma^\uparrow$, $\sigma^\downarrow$ and $r_\mathrm{lim}$ are. Since the center-of-mass position is extracted from the measured asymmetry as $\widehat{y} = \dPhat / s$, we look at the uncertainty of the inverse slope factor $s^{-1}_\mathrm{opt}$.

Simultaneously varying $\sigma^\uparrow$ (or $\sigma^\downarrow$) by $25\%$ and the limiting radius $r_\mathrm{lim}$ from $\un{3}{mm}$ to the full cutoff-tube radius of $\un{5}{mm}$ changes the calculated value of $s^{-1}_\mathrm{opt}$ by a maximum of $8.5\%$. We will use this number as a conservative estimate of the uncertainty of the inverse slope factor due to the density distribution of atoms in the cavity.
This result once again assumes a numerical solution of the differential equations. Calculating the effective slope values from \eref{eq_dP_definition} directly would introduce an additional error of around $6\%$ due to the non-depletion approximation.

A comparable uncertainty contribution arises due to the use of the approximated analytic field distribution of \eref{eq_amp_Hrho} and \eref{eq_amp_Hz}: The radial variation of the tipping angle for the actual fields in a cavity with endcap apertures is~\cite{Li2010}
\begin{equation}
\label{eq_k_rho}
\func{\Theta_\mathrm{act}}{\rho} = \frac{\pi}{2} \, \eta \, b \, \func{J_0}{k \, \rho}
\ ,
\end{equation}
where $k=\omega / c$ is the wave vector of the applied frequency. The factor $\eta$ is a constant that depends only on the cutoff tube radius and is equivalent to $\zeta$ in \eref{eq_Omega_x}-\eref{eq_Omega_z}. For CSF2, the expression for $\Theta$ derived from the approximate fields and given in \eref{eq_tipping_angle_gamma_rho} underestimates the tipping angle by $3.6\%$ on the cavity axis and overestimates it by $4.8\%$ at the cutoff tube radius. This provides a measure of the components of the actual $H_z$ fields that arise due to the presence of the endcap apertures and are neglected in the analytical model. For $H_\rho$, the neglected components can be assumed to be of similar magnitude. Since the asymmetry measurements are taken near the extrema of the Fourier transform $\widetilde{H}_\rho$ (see \sref{sec_measurement_detuning}), it is unlikely that the neglected field components have over-proportionally large Fourier components at these frequencies.

Realizing that $\dP$ is essentially given by the product of $\widetilde{H}_\rho$ and $\widetilde{H}_z$  (which provide the $\Xfunc$ and $\Yfunc$ terms of \eref{eq_linear_combination_with_tilt}) then gives a maximum expected increase of the asymmetry by $(1.048)^2$. We will represent this as an assumed $10\%$ uncertainty of the inverse slope factor. The value depends on the aspect ratio of the cavity, however: While the $\gamma$ term in \eref{eq_tipping_angle_gamma_rho} varies with the cavity radius, $k$ in \eref{eq_k_rho} does not.

The final significant uncertainty contribution results from the uncertainty of the parameters used to calculate the effective slope factor. This can be estimated from \eref{eq_dP_equation}: Approximating the term in brackets as 1 and combining the result with \eref{eq_tilt_from_Pres}, this yields
\begin{equation}
\label{eq_simplified_slope_from_P0}
 s_\mathrm{simp}
  = \sqrt{c_{01}} \frac{ \left(b \, \zeta \right)^2 }{ b_\mathrm{nom} }
    \frac{\pi^4}{64 \, d} \,
    \frac{ \Xfunc \, \Yfunc }{ \cos \alpha }
    \sqrt{ \widehat{P}_{01}^\mathrm{res}}
    \ .
\end{equation}
Standard uncertainty propagation can now be applied. For simplicity we assume $\cos \alpha = 1$ and parameter values typical for CSF2. A predicted value for $\widehat{P}^\mathrm{res}_{01}$ at optimized parameters is calculated from \eref{eq_P01_res}.
Significant uncertainty contributions arise from the measurement of $\widehat{P}$, with an assumed uncertainty of $\sigma_\mathrm{P} = 5 \times 10^{-4}$ (see \sref{sec_experimental_test}), and from the uncertainty of the \rf amplitude $\sigma_\mathrm{b}=0.1$. In total, the calculation yields a relative uncertainty of the inverse slope factor of $6.4\%$.

Adding all three contributions in quadrature, we assume a total relative uncertainty of $u_s = 15\%$ for the inverse slope factor at optimized parameters.

Similar arguments as in \sref{sec_centering_uncertainty} can be made for the part of the position uncertainty that does not depend on the slope factor. Assuming the same values for $\sigma_{y}^\mathrm{sing}$ and $\sigma_{y}^\mathrm{dist}$ as before, the total uncertainty is then given by
\begin{equation}
\sigma_y^\mathrm{tot} 
= \sqrt{ 
      \left(\sigma_{y}^\mathrm{sing} \right)^2
    + \left(\sigma_{y}^\mathrm{dist} \right)^2
    + \left( u_s \  s^{-1}_\mathrm{opt} \  \dPhat  \right)^2
  }
  \ .
\end{equation}
This yields $\sigma_y^\mathrm{tot} = \un{0.14}{mm}$ for a cloud with a center-of-mass position of $\widehat{y}=\un{0.5}{mm}$, as encountered during the upwards passage in CSF2.

\section{Summary}
\label{sec_summary}

We have demonstrated a method to obtain information on the position of atoms in the cavity of a fountain clock by applying a slight tilt of the normally vertical C-field. Since measurements are performed using the actual field in the cavity and the actual detection system, the results represent the position of the subset of atoms that contribute to frequency measurements. 


Although the C-field tilt in the fountain CSF2 was originally unintended, the application of this method provided valuable information on the cloud positions during the two cavity passages, which has already been used in a new DCP evaluation that was recently published~\cite{Weyers2012}.

Systematic position measurements will be possible for any fountain clock that is equipped with a means to provide a controlled horizontal component of the magnetic field on the order of $\un{10}{nT}$ in the cavity region. The methods shown here then allow for efficient measurements of the cloud position that might even be periodically performed between frequency measurements. Since the results also provide information on the precise value of the C-field tilt, they can also be used to exclude an unwanted tilt during the actual frequency measurements.

\begin{acknowledgments}

The authors thank K. Gibble for extensive discussions and valuable suggestions on all aspects of this work.
This work was supported by Deutsche Forschungsgemeinschaft.

\end{acknowledgments}


\appendix


\section{Quantum mechanical derivation of the transition probability}
\label{sec_QM_derivation}

Following a standard textbook approach, 
we start from a two-level model with the eigenstates $\ket{\psi_0}$ and $\ket{\psi_1}$, such that the state as a function of time can be written as the linear combination
\begin{equation}
\ket{\func{\Psi}{t}} = \func{c_0}{t} \ket{\psi_0} + \func{c_1}{t} \ket{\psi_1} \ .
\end{equation}
The coefficients $c_0$ and $c_1$ are generally complex and incorporate information on the phase of the atomic state. The behavior of the system in the presence of an oscillating field can, after applying the rotating wave approximation, be described by the two coupled differential equations
\begin{eqnarray}
\label{eq_DE1}
\func{c'_0}{t} 
&=& - i \, \tfrac{1}{2} \,
    \func{\Omega^*}{t} \, 
    \func{\exp}{ i \, \Delom \, t } \func{c_1}{t} \\
\label{eq_DE2}
\func{c'_1}{t} 
&=& - i \, \tfrac{1}{2} \, 
    \func{\Omega}{t} \,
    \func{\exp}{- i \, \Delom \, t} \func{c_{0}}{t} 
\ .
\end{eqnarray}

Here $\Delom = \omega - \omega_\mathrm{res}$ is the detuning of the microwave signal from the atomic resonance. The \rf field amplitude and phase $\varphi$ are described in terms of the complex Rabi frequency $\Omega$ (see \sref{sec_complex_Rabi}). An analytic solution exists for the case of a resonant pulse with constant phase: For an initial state described by $\func{c_0}{t_i} = 1$ and $\func{c_1}{t_i} = 0$, the transition probability depends on
\begin{equation}
  \label{eq_tipping_angle}  
  \func{c_1}{t_f} = -i \, \sin \tfrac{1}{2} \Theta  \ ,
  \quad \mathrm{with} \
  \Theta = \left| \int_{t_i}^{t_f} \func{\Omega}{t} dt \right|
  \ ,
\end{equation}
where $\Theta$ will be referred to here as the resonant tipping angle. The probability $P$ of finding the atom in the excited state at $t_f$, after the end of the pulse is then
\begin{equation}
\label{eq_P_from_Theta}
P 
= \abs{ \func{c_1}{t_f} }^2 
= \left( 1 - \cos \Theta \right) / 2
\ .
\end{equation}

Insight into the off-resonant transition probabilities can be gained by using a non-depletion approximation: For small transition probabilities $c_0$ is approximately 1, so that \eref{eq_DE2} can be integrated to give
\begin{equation}
 \func{c_1^\mathrm{nda}}{t_f} = - i \, \tfrac{1}{2} \int_{t_i}^{t_f} \func{\Omega}{t} \func{\exp}{- i \, \Delom \, t} dt
 \ .
\end{equation}
By letting $t_i$ and $t_f$ go to $-\infty$ and $\infty$ respectively, the integral becomes identical to the Fourier transform of $\func{\Omega}{t}$, where we use the convention
\begin{equation}
\label{eq_Fourier_transform}
\func{\widetilde\Omega}{\Delom} = \int_{-\infty}^{\infty} \func{\Omega}{t} \func{\exp}{ - i \, \Delom \, t} dt
\ ,
\end{equation}
because it directly includes the resonant tipping angle as $\Theta=\left| \func{\widetilde{\Omega}}{0} \right|$, which is valid even outside the non-depletion approximation.

\section{Derivation of the tilt angle equation}
\label{sec_tilt_derivation}

Averaged over the radial distribution $\func{W_r}{\rho}$ introduced in \eref{eq_P00_avg}, the resonant \trans{0}{-1} transition probability given by \eref{eq_P01_res} becomes
\begin{equation}
\widehat{P}_{01}^\mathrm{res}  
= \tfrac{1}{4} \, c_{01} 
  \int_0^{r_a}  
    \left(    b \, \zeta \,
      \tfrac{\pi}{2} \, \func{J_0}{\gamma \rho } \, \tan \alpha 
    \right)^2
    \func{W_r}{\rho}
    \ d\rho
    \ .
\end{equation}
Using $\left(\func{J_0}{x} \, \right)^2 = 2 \func{J_0}{x} -1 + \func{O}{x}^4$, this yields
\begin{eqnarray}
\widehat{P}_{01}^\mathrm{res}  
= \tfrac{1}{4} \, c_{01}
  \left(b \, \zeta \, \tfrac{\pi}{2} \tan \alpha \right)^2
  \Bigg(
&&  2 \int_0^{r_a} 
       \func{J_0}{\gamma \rho }
       \func{W_r}{\rho} 
       \, d\rho
\nonumber \\
&&  - \int_0^{r_a} \func{W_r}{\rho}
       \, d\rho  
  \Bigg)
\ ,
\end{eqnarray}
where the small fourth order term has been ignored. The normalization of $\func{W_r}{\rho}$ makes the second integral $1$. And since the same distribution was used to optimize the \rf amplitude, the approximation 
\begin{eqnarray}
   \tfrac{\pi}{2} \, b_\mathrm{nom}
  &\approx & \int_0^{r_a}       
    \func{\Theta}{\rho}
    \func{W_r}{\rho}
    \, d\rho  
\nonumber \\
  & = &\int_0^{r_a}       
    b \, \zeta \tfrac{\pi}{2} \, \func{J_0}{\gamma \rho } \,
    \func{W_r}{\rho}
    \, d\rho      
\label{eq_bnom_approximation}
\end{eqnarray}
%
can be used to replace the first integral and find
\begin{equation}
\widehat{P}_{01}^\mathrm{res}  
= \tfrac{1}{4} \, c_{01}
  \left(\tfrac{\pi}{2} \, \tan \alpha \right)^2
  \left( b_\mathrm{nom} + \Delta b \right)
  \left( b_\mathrm{nom} - \Delta b \right)  
\ .
\end{equation}
Since $\Delta b = b \, \zeta - b_\mathrm{nom}$ is small, this gives
\begin{equation}
\label{eq_P01_avg}
\widehat{P}_{01}^\mathrm{res}  
= \tfrac{1}{4} \, c_{01}
  \left(\tfrac{\pi}{2} \, b_\mathrm{nom} \, \tan \alpha \right)^2
  \ .
\end{equation}


\begin{thebibliography}{17}%
\makeatletter
\providecommand \@ifxundefined [1]{%
 \@ifx{#1\undefined}
}%
\providecommand \@ifnum [1]{%
 \ifnum #1\expandafter \@firstoftwo
 \else \expandafter \@secondoftwo
 \fi
}%
\providecommand \@ifx [1]{%
 \ifx #1\expandafter \@firstoftwo
 \else \expandafter \@secondoftwo
 \fi
}%
\providecommand \natexlab [1]{#1}%
\providecommand \enquote  [1]{``#1''}%
\providecommand \bibnamefont  [1]{#1}%
\providecommand \bibfnamefont [1]{#1}%
\providecommand \citenamefont [1]{#1}%
\providecommand \href@noop [0]{\@secondoftwo}%
\providecommand \href [0]{\begingroup \@sanitize@url \@href}%
\providecommand \@href[1]{\@@startlink{#1}\@@href}%
\providecommand \@@href[1]{\endgroup#1\@@endlink}%
\providecommand \@sanitize@url [0]{\catcode `\\12\catcode `\$12\catcode
  `\&12\catcode `\#12\catcode `\^12\catcode `\_12\catcode `\%12\relax}%
\providecommand \@@startlink[1]{}%
\providecommand \@@endlink[0]{}%
\providecommand \url  [0]{\begingroup\@sanitize@url \@url }%
\providecommand \@url [1]{\endgroup\@href {#1}{\urlprefix }}%
\providecommand \urlprefix  [0]{URL }%
\providecommand \Eprint [0]{\href }%
\providecommand \doibase [0]{http://dx.doi.org/}%
\providecommand \selectlanguage [0]{\@gobble}%
\providecommand \bibinfo  [0]{\@secondoftwo}%
\providecommand \bibfield  [0]{\@secondoftwo}%
\providecommand \translation [1]{[#1]}%
\providecommand \BibitemOpen [0]{}%
\providecommand \bibitemStop [0]{}%
\providecommand \bibitemNoStop [0]{.\EOS\space}%
\providecommand \EOS [0]{\spacefactor3000\relax}%
\providecommand \BibitemShut  [1]{\csname bibitem#1\endcsname}%
\let\auto@bib@innerbib\@empty
\bibitem [{\citenamefont {Wynands}\ and\ \citenamefont
  {Weyers}(2005)}]{Wynands2005}%
  \BibitemOpen
  \bibfield  {author} {\bibinfo {author} {\bibfnamefont {R.}~\bibnamefont
  {Wynands}}\ and\ \bibinfo {author} {\bibfnamefont {S.}~\bibnamefont
  {Weyers}},\ }\href@noop {} {\bibfield  {journal} {\bibinfo  {journal}
  {Metrologia}\ }\textbf {\bibinfo {volume} {42}},\ \bibinfo {pages} {S64}
  (\bibinfo {year} {2005})}\BibitemShut {NoStop}%
\bibitem [{\citenamefont {Li}\ and\ \citenamefont {Gibble}(2004)}]{Li2004}%
  \BibitemOpen
  \bibfield  {author} {\bibinfo {author} {\bibfnamefont {R.}~\bibnamefont
  {Li}}\ and\ \bibinfo {author} {\bibfnamefont {K.}~\bibnamefont {Gibble}},\
  }\href@noop {} {\bibfield  {journal} {\bibinfo  {journal} {Metrologia}\
  }\textbf {\bibinfo {volume} {41}},\ \bibinfo {pages} {376} (\bibinfo {year}
  {2004})}\BibitemShut {NoStop}%
\bibitem [{\citenamefont {Li}\ and\ \citenamefont {Gibble}(2010)}]{Li2010}%
  \BibitemOpen
  \bibfield  {author} {\bibinfo {author} {\bibfnamefont {R.}~\bibnamefont
  {Li}}\ and\ \bibinfo {author} {\bibfnamefont {K.}~\bibnamefont {Gibble}},\
  }\href@noop {} {\bibfield  {journal} {\bibinfo  {journal} {Metrologia}\
  }\textbf {\bibinfo {volume} {47}},\ \bibinfo {pages} {534} (\bibinfo {year}
  {2010})}\BibitemShut {NoStop}%
\bibitem [{\citenamefont {Gu\'ena}\ \emph {et~al.}(2011)\citenamefont
  {Gu\'ena}, \citenamefont {Li}, \citenamefont {Gibble}, \citenamefont {Bize},\
  and\ \citenamefont {Clairon}}]{Guena2011PRL}%
  \BibitemOpen
  \bibfield  {author} {\bibinfo {author} {\bibfnamefont {J.}~\bibnamefont
  {Gu\'ena}}, \bibinfo {author} {\bibfnamefont {R.}~\bibnamefont {Li}},
  \bibinfo {author} {\bibfnamefont {K.}~\bibnamefont {Gibble}}, \bibinfo
  {author} {\bibfnamefont {S.}~\bibnamefont {Bize}}, \ and\ \bibinfo {author}
  {\bibfnamefont {A.}~\bibnamefont {Clairon}},\ }\href@noop {} {\bibfield
  {journal} {\bibinfo  {journal} {Phys. Rev. Lett.}\ }\textbf {\bibinfo
  {volume} {106}},\ \bibinfo {pages} {130801} (\bibinfo {year}
  {2011})}\BibitemShut {NoStop}%
\bibitem [{\citenamefont {Li}\ \emph {et~al.}(2011)\citenamefont {Li},
  \citenamefont {Gibble},\ and\ \citenamefont {Szymaniec}}]{Szymaniec2011}%
  \BibitemOpen
  \bibfield  {author} {\bibinfo {author} {\bibfnamefont {R.}~\bibnamefont
  {Li}}, \bibinfo {author} {\bibfnamefont {K.}~\bibnamefont {Gibble}}, \ and\
  \bibinfo {author} {\bibfnamefont {K.}~\bibnamefont {Szymaniec}},\ }\href@noop
  {} {\bibfield  {journal} {\bibinfo  {journal} {Metrologia}\ }\textbf
  {\bibinfo {volume} {48}},\ \bibinfo {pages} {283} (\bibinfo {year}
  {2011})}\BibitemShut {NoStop}%
\bibitem [{\citenamefont {Weyers}\ \emph {et~al.}(2012)\citenamefont {Weyers},
  \citenamefont {Gerginov}, \citenamefont {Nemitz}, \citenamefont {Li},\ and\
  \citenamefont {Gibble}}]{Weyers2012}%
  \BibitemOpen
  \bibfield  {author} {\bibinfo {author} {\bibfnamefont {S.}~\bibnamefont
  {Weyers}}, \bibinfo {author} {\bibfnamefont {V.}~\bibnamefont {Gerginov}},
  \bibinfo {author} {\bibfnamefont {N.}~\bibnamefont {Nemitz}}, \bibinfo
  {author} {\bibfnamefont {R.}~\bibnamefont {Li}}, \ and\ \bibinfo {author}
  {\bibfnamefont {K.}~\bibnamefont {Gibble}},\ }\href@noop {} {\bibfield
  {journal} {\bibinfo  {journal} {Metrologia}\ }\textbf {\bibinfo {volume}
  {49}},\ \bibinfo {pages} {82} (\bibinfo {year} {2012})}\BibitemShut {NoStop}%
\bibitem [{\citenamefont {Schr\"oder}\ \emph {et~al.}(2002)\citenamefont
  {Schr\"oder}, \citenamefont {H\"ubner},\ and\ \citenamefont
  {Griebsch}}]{Schroeder2002}%
  \BibitemOpen
  \bibfield  {author} {\bibinfo {author} {\bibfnamefont {R.}~\bibnamefont
  {Schr\"oder}}, \bibinfo {author} {\bibfnamefont {U.}~\bibnamefont
  {H\"ubner}}, \ and\ \bibinfo {author} {\bibfnamefont {D.}~\bibnamefont
  {Griebsch}},\ }\href@noop {} {\bibfield  {journal} {\bibinfo  {journal} {IEEE
  Transactions on Ultrasonics, Ferroelectrics and Frequency Control}\ }\textbf
  {\bibinfo {volume} {49}},\ \bibinfo {pages} {383} (\bibinfo {year}
  {2002})}\BibitemShut {NoStop}%
\bibitem [{\citenamefont {Gerginov}\ \emph {et~al.}(2010)\citenamefont
  {Gerginov}, \citenamefont {Nemitz}, \citenamefont {Weyers}, \citenamefont
  {Schr\"oder}, \citenamefont {Griebsch},\ and\ \citenamefont
  {Wynands}}]{Gerginov2010}%
  \BibitemOpen
  \bibfield  {author} {\bibinfo {author} {\bibfnamefont {V.}~\bibnamefont
  {Gerginov}}, \bibinfo {author} {\bibfnamefont {N.}~\bibnamefont {Nemitz}},
  \bibinfo {author} {\bibfnamefont {S.}~\bibnamefont {Weyers}}, \bibinfo
  {author} {\bibfnamefont {R.}~\bibnamefont {Schr\"oder}}, \bibinfo {author}
  {\bibfnamefont {D.}~\bibnamefont {Griebsch}}, \ and\ \bibinfo {author}
  {\bibfnamefont {R.}~\bibnamefont {Wynands}},\ }\href
  {http://stacks.iop.org/0026-1394/47/i=1/a=008} {\bibfield  {journal}
  {\bibinfo  {journal} {Metrologia}\ }\textbf {\bibinfo {volume} {47}},\
  \bibinfo {pages} {65} (\bibinfo {year} {2010})}\BibitemShut {NoStop}%
\bibitem [{\citenamefont {Weyers}\ \emph {et~al.}(2006)\citenamefont {Weyers},
  \citenamefont {Schr\"oder},\ and\ \citenamefont {Wynands}}]{Weyers2006}%
  \BibitemOpen
  \bibfield  {author} {\bibinfo {author} {\bibfnamefont {S.}~\bibnamefont
  {Weyers}}, \bibinfo {author} {\bibfnamefont {R.}~\bibnamefont {Schr\"oder}},
  \ and\ \bibinfo {author} {\bibfnamefont {R.}~\bibnamefont {Wynands}},\ }in\
  \href@noop {} {\emph {\bibinfo {booktitle} {Proc. 20th European Frequency and
  Time Forum (Braunschweig, Germany)}}}\ (\bibinfo {year} {2006})\ pp.\
  \bibinfo {pages} {173--–80}\BibitemShut {NoStop}%
\bibitem [{\citenamefont {Shirley}\ \emph {et~al.}(2006)\citenamefont
  {Shirley}, \citenamefont {Levi}, \citenamefont {Heavner}, \citenamefont
  {Calonico}, \citenamefont {Yu},\ and\ \citenamefont
  {Jefferts}}]{Shirley2006}%
  \BibitemOpen
  \bibfield  {author} {\bibinfo {author} {\bibfnamefont {J.~H.}\ \bibnamefont
  {Shirley}}, \bibinfo {author} {\bibfnamefont {F.}~\bibnamefont {Levi}},
  \bibinfo {author} {\bibfnamefont {T.~P.}\ \bibnamefont {Heavner}}, \bibinfo
  {author} {\bibfnamefont {D.}~\bibnamefont {Calonico}}, \bibinfo {author}
  {\bibfnamefont {D.-H.}\ \bibnamefont {Yu}}, \ and\ \bibinfo {author}
  {\bibfnamefont {S.}~\bibnamefont {Jefferts}},\ }\href@noop {} {\bibfield
  {journal} {\bibinfo  {journal} {IEEE Trans. Ultrasonics, Ferroelectrics and
  Frequency Control}\ }\textbf {\bibinfo {volume} {53}},\ \bibinfo {pages}
  {2376} (\bibinfo {year} {2006})}\BibitemShut {NoStop}%
\bibitem [{\citenamefont {Foot}(2005)}]{Foot2005}%
  \BibitemOpen
  \bibfield  {author} {\bibinfo {author} {\bibfnamefont {C.~J.}\ \bibnamefont
  {Foot}},\ }\href@noop {} {\emph {\bibinfo {title} {Atomic Physics}}}\
  (\bibinfo  {publisher} {New York: Oxford University Press},\ \bibinfo {year}
  {2005})\ p.\ \bibinfo {pages} {15ff}\BibitemShut {NoStop}%
\bibitem [{\citenamefont {Millman}(1939)}]{Millman1939}%
  \BibitemOpen
  \bibfield  {author} {\bibinfo {author} {\bibfnamefont {S.}~\bibnamefont
  {Millman}},\ }\href@noop {} {\bibfield  {journal} {\bibinfo  {journal} {Phys.
  Rev.}\ }\textbf {\bibinfo {volume} {55}},\ \bibinfo {pages} {628} (\bibinfo
  {year} {1939})}\BibitemShut {NoStop}%
\bibitem [{\citenamefont {Vanier}\ \emph {et~al.}(1984)\citenamefont {Vanier},
  \citenamefont {Mungall},\ and\ \citenamefont {Boulanger}}]{Vanier1984}%
  \BibitemOpen
  \bibfield  {author} {\bibinfo {author} {\bibfnamefont {J.}~\bibnamefont
  {Vanier}}, \bibinfo {author} {\bibfnamefont {A.~G.}\ \bibnamefont {Mungall}},
  \ and\ \bibinfo {author} {\bibfnamefont {J.-S.}\ \bibnamefont {Boulanger}},\
  }\href@noop {} {\bibfield  {journal} {\bibinfo  {journal} {Metrologia}\
  }\textbf {\bibinfo {volume} {20}},\ \bibinfo {pages} {101} (\bibinfo {year}
  {1984})}\BibitemShut {NoStop}%
\bibitem [{\citenamefont {Giordano}\ \emph {et~al.}(1995)\citenamefont
  {Giordano}, \citenamefont {Pichon}, \citenamefont {C\'erez},\ and\
  \citenamefont {Th\'eobald}}]{Giordano1995}%
  \BibitemOpen
  \bibfield  {author} {\bibinfo {author} {\bibfnamefont {V.}~\bibnamefont
  {Giordano}}, \bibinfo {author} {\bibfnamefont {L.}~\bibnamefont {Pichon}},
  \bibinfo {author} {\bibfnamefont {P.}~\bibnamefont {C\'erez}}, \ and\
  \bibinfo {author} {\bibfnamefont {G.}~\bibnamefont {Th\'eobald}},\
  }\href@noop {} {\bibfield  {journal} {\bibinfo  {journal} {J. Appl. Phys.}\
  }\textbf {\bibinfo {volume} {78}},\ \bibinfo {pages} {1} (\bibinfo {year}
  {1995})}\BibitemShut {NoStop}%
\bibitem [{\citenamefont {Vanier}\ and\ \citenamefont
  {Audoin}(1989)}]{Vanier1989}%
  \BibitemOpen
  \bibfield  {author} {\bibinfo {author} {\bibfnamefont {J.}~\bibnamefont
  {Vanier}}\ and\ \bibinfo {author} {\bibfnamefont {C.}~\bibnamefont
  {Audoin}},\ }\href@noop {} {\emph {\bibinfo {title} {The Quantum Physics of
  Atomic Frequency Standards}}}\ (\bibinfo  {publisher} {Bristol: Hilger},\
  \bibinfo {year} {1989})\BibitemShut {NoStop}%
\bibitem [{Note1()}]{Note1}%
  \BibitemOpen
  \bibinfo {note} {The transition strength coefficient for \unhbox \voidb@x
  \hbox {[-1 to 0]} is $c_{10}=3/16$.}\BibitemShut {Stop}%
\bibitem [{\citenamefont {Breit}\ and\ \citenamefont {Rabi}(1931)}]{Breit1931}%
  \BibitemOpen
  \bibfield  {author} {\bibinfo {author} {\bibfnamefont {G.}~\bibnamefont
  {Breit}}\ and\ \bibinfo {author} {\bibfnamefont {I.~I.}\ \bibnamefont
  {Rabi}},\ }\href@noop {} {\bibfield  {journal} {\bibinfo  {journal} {Phys.
  Rev.}\ }\textbf {\bibinfo {volume} {38}},\ \bibinfo {pages} {2082} (\bibinfo
  {year} {1931})}\BibitemShut {NoStop}%
\end{thebibliography}
\end{document}